\documentclass[10pt,twocolumn,twoside,letterpaper]{IEEEtran}

\usepackage{amsmath}
\usepackage{amssymb}
\usepackage{amsthm}
\usepackage{bm}
\usepackage{algorithm}
\usepackage{algpseudocode}
\usepackage[nocompress]{cite}
\usepackage{enumerate}
\usepackage{mathrsfs}
\usepackage{ifthen}
\usepackage[T1]{fontenc}
\usepackage{url}
\usepackage{txfonts}
\usepackage{comment}

\DeclareMathOperator{\cir}{cir}
\DeclareMathOperator{\lyr}{lyr}
\DeclareMathOperator{\wt}{wt}

\newcommand{\F}{\mathbb{F}}

\newcommand{\R}{\mathscr{R}}
\newcommand{\Z}{\mathbb{Z}}

\newcommand{\Input}{\item[\textsc{Input:}]}
\newcommand{\Output}{\item[\textsc{Output:}]}
\newcommand{\Remark}{\item[\textsc{Remark:}]}
\newcommand{\ZComment}[1]{\(\triangleright\) #1}
\newcommand{\LComment}[1]{\item[\(\triangleright\)] #1}
\renewcommand{\And}{\; \textbf{and} \;}
\newcommand{\Or}{\; \textbf{or} \;}
\newcommand{\To}{\; \textbf{to} \;}
\newcommand{\SuchThat}{\textbf{such that}}
\newcommand{\tp}{\textrm{\tiny T}}

\newtheorem{definition}{Definition}

\newboolean{anonymous}
\setboolean{anonymous}{false}

\begin{document}

\title{\bf Scaling efficient code-based cryptosystems for embedded platforms}

\ifthenelse{\boolean{anonymous}}
{ 
\author{
    Anonymized for submission
}
} 
{ 
\author{
    Felipe P. Biasi,
    Paulo~S.~L.~M.~Barreto$^*$\thanks{$^*$ P. Barreto is the corresponding author: \texttt{pbarreto@larc.usp.br}},
    Rafael~Misoczki,
    Wilson~V.~Ruggiero
\thanks{
    P. Barreto is supported by the Brazilian National Council for Scientific and Technological Development (CNPq) under research productivity grant 303163/2009-7.
}
\thanks{
    F. Biasi, P. Barreto and W. Ruggiero are supported by Cisco Research Award 2011-050 `Alternate Public Key Cryptosystems'
}
\thanks{
    F. Biasi, W. Ruggiero and P. Barreto are with the Department of Computer and Digital Systems Engineering,
    Escola Polit\'{e}cnica, Universidade de S\~{a}o Paulo, Brazil.
    (e-mail: \texttt{\{fbiasi,pbarreto,wilson\}@larc.usp.br})
}
\thanks{
    R. Misoczki is with SECRET team, INRIA Paris-Rocquencourt, France.
    (e-mail: \texttt{rafael.misoczki@inria.fr})
}
}
} 

\markboth{
}{Biasi \MakeLowercase{\textit{et al.}}: Scaling efficient syndrome-based cryptosystems for embedded platforms}

\maketitle
\IEEEpeerreviewmaketitle

\begin{abstract}
We describe a family of highly efficient codes for cryptographic purposes and dedicated algorithms for their manipulation. Our proposal is especially tailored for highly constrained platforms, and surpasses certain conventional and post-quantum proposals (like RSA and NTRU, respectively) according to most if not all efficiency metrics.
\end{abstract}



\section{Introduction}\label{sec:intro}

\IEEEPARstart{Q}{uantum} computers, should they become a technological reality, will pose a threat to public-key cryptosystems based on certain intractability assumptions, like the integer factorization problem, or IFP (like RSA), and the discrete logarithm problem, or DLP (like Diffie-Hellman or DSA, in their elliptic curve version or otherwise). To face this scenario, several cryptosystems have been proposed that apparently resist attacks mounted with the help of quantum computers. The security of these so-called post-quantum cryptosystems~\cite{bernstein-buchmann-dahmen} stems from quite distinct computational intractability assumptions. Such schemes are not necessarily new --- for instance, cryptosystems based on coding theory (specifically, on the intractability of the syndrome decoding problem, or SDP) are known for nearly as long as the very concept of asymmetric cryptography itself, though they have only recently been attracting renewed interest.

However, being quantum-resistant is not the only interesting feature of many post-quantum proposals --- some of them are equally remarkable because of their improved efficiency and simplicity for certain types of applications relatively to conventional schemes. Thus, schemes based on the SDP entirely avoid the multiprecision integer arithmetic typically needed by IFP or DLP cryptosystems, and their computational cost is usually a few orders of complexity smaller than those systems, reaching $\tilde{O}(n)$ instead of $\tilde{O}(n^2)$ or $\tilde{O}(n^3)$ which are commonplace in pre-quantum schemes. This indicates that post-quantum alternatives may have advantages even in situations where quantum attacks are not the main concern, and justifies the investigation on how advantageous they can be.

Particularly interesting scenarios where such post-quantum schemes may have a positive impact are wireless sensor networks~\cite{aysal-barner,oliveira-barros} and the so-called ``Internet of Things,'' in which a wide range of devices are interconnected, from the most powerful clustered servers to embedded systems with extremely limited processing resources, storage, bandwidth occupation and power consumption, including microcontrollers~\cite{koschuch-lechner-weitzer-grossschaedl-szekely-tillich-wolkerstorfer,vogt-poschmann-paar} and dedicated hardware~\cite{maes-schellekens-verbauwhede}.

One of the leading families of post-quantum cryptographic schemes is that of code-based cryptosystems~\cite{mceliece,niederreiter}. In contrast with the form in which these systems were originally proposed, where key sizes were typically large, modern approaches do offer far more space-efficient parameters, being fairly practical on general-purpose platforms. On can thus ask whether such schemes are suitable for highly constrained platforms as well.

Low-density parity-check (LDPC) codes and their quasi-cyclic variants (QC-LDPC) have been proposed for cryptographic applications
\cite{monico-rosenthal-shokrollahi,baldi-chiaraluce-garello,baldi-chiaraluce-garello-mininni,baldi-chiaraluce,baldi-bodrato-chiaraluce,baldi-bianchi-chiaraluce,klinc-jeongseok-mclaughlin-barros-byungjae,thangaraj-dihidar-calderbank-mclaughlin-merolla} although in a form still unsuitable for constrained platforms.
Recently, quasi-cyclic moderate-density parity-check (QC-MDPC) codes have been designed to provide strong security assurances for McEliece-style cryptosystems~\cite{misoczki-sendrier-tillich-barreto}. Such codes are arguably ideal for modern general-purpose platforms, matching or surpassing the processing efficiency of conventional cryptosystems. However, no assessment of their suitability for constrained platform has been made, and indeed the traditional bit-flipping and belief-propagation decoding methods, even though they are quite processing-efficient, appear at first glance unsuitable for an Internet-of-Things scenario due to their considerable storage requirements.

\subsection{Our Results}\label{sec:our-contrib}

Our contributions in this paper are twofold:
\begin{itemize}
\item On the one hand, a family of linear error-correcting codes (so-called CS-MDPC codes) that are highly efficient for cryptographic applications in terms of reduced per-key and per-message bandwidth occupation;
\item On the other hand, an efficient decoder for that family of codes that is especially tailored for (though not restricted to) constrained platforms.
\end{itemize}
Specifically, we show how to obtain code-based cryptosystems where the public keys and the space overhead incurred for each cryptogram are comparable in size to, or even smaller than, the corresponding values for the RSA cryptosystem at practical security levels. A careful selection of design features for the key generation, encoding, and decoding algorithms lead to very short processing times, and executable code size in software or area occupation in hardware (and thus potentially also energy consumption) tend to be considerably smaller than what can be attained with RSA or elliptic curve cryptosystems. Our proposed variant of the bit flipping decoding technique needs only $O(1)$ ancillary storage, in comparison with $O(n)$ (where $n$ in the code length) as in previous variants of that technique.

\subsection{Organization of the Paper}

The remainder of this document is organized as follows.
We provide theoretical preliminaries in Section~\ref{sec:prelim}, including LDPC and MDPC codes, the hard decision decoding method, and code-based cryptosystems.
We describe the new family of codes and assess its security properties in Section~\ref{sec:cs-mdpc}.
In Section~\ref{sec:scaling} we outline our proposed techniques to deploy code-based cryptosystems on embedded platforms, in particular an efficient bit-flipping decoder that takes only $O(1)$ ancillary storage instead of the usual $O(n)$ requirements.
We illustrate some suggested parameters for typical security levels in Section~\ref{sec:params} and assess the overall results of our proposal experimentally in Section~\ref{sec:experim}.
We conclude in Section~\ref{sec:conclusion}.

\section{Preliminaries}\label{sec:prelim}

\subsection{General notation}

Matrix and vector indices will be numbered from 0 throughout this paper, unless otherwise stated. Let $p$ be a prime and let $q = p^m$ for some $m > 0$. The finite field of $q$ elements is written $\F_q$. Given $h \in \F_2^r$, we denote by $\cir(h)$ the circulant matrix
\[
\cir(h) = \left[
\begin{array}{cccc}
h_0     & h_1    & \dots  & h_{r-1}\\
h_{r-1} & h_0    & \dots  & h_{r-2}\\
\vdots  & \vdots & \ddots & \vdots\\
h_1     & h_2    & \dots  & h_0
\end{array}
\right].
\]

\subsection{Error Correcting Codes}

A (binary) linear $[n, k]$ error-correcting code $\mathscr{C}$ is a subspace of $\F_2^n$ of dimension $k$. Such a code is specified by either a \emph{generator matrix} $G \in \F_2^{k \times n}$ such that $\mathscr{C} = \{u G \in \F_2^n \mid u \in \F_2^k\}$, or else by a \emph{parity-check matrix} $H \in \F_2^{r \times n}$ such that $\mathscr{C} = \{v \in \F_2^n \mid v H^{\tp} = 0^r\}$ where $r = n - k$.

We will be particularly interested in \emph{quasi-cyclic} codes, namely, codes that admit a parity-check matrix consisting of $n_0$ horizontally joined circulant square blocks of size $r \times r$. Thus:
\[
H = [\cir(h_0) \mid \cir(h_1) \mid \dots \mid \cir(h_{n_0-1})],
\]
where $h_i \in \F_2^r$, $0 \leqslant i < n_0$. The representation advantages of such codes are obvious, since $H$ can be compactly stored as $n_0$ sequences of $r$ bits each.

The \emph{syndrome decoding problem} (SDP) consists of computing an error pattern $e \in \F_2^n$ given a parity-check matrix $H \in \F_2^{r \times n}$ for the underlying code, and a syndrome $s = e H^{\tp} \in \F_2^r$. In general the SDP is NP-hard, but sometimes the knowledge of certain structural code properties makes this problem solvable in polynomial time.

\subsection{LDPC codes}\label{sec:ldpc}

LDPC codes were invented by Robert Gallager~\cite{gallager} and are linear codes obtained from sparse bipartite graphs. Suppose that $\mathscr{G}$ is a graph with $n$ left nodes (called message nodes) and $r$ right nodes (called check nodes). The graph gives rise to a linear code of block length $n$ and dimension at least $n - r$ in the following way: The $n$ coordinates of the codewords are associated with the $n$ message nodes. The codewords are those vectors $(c_1, \dots, c_n)$ such that for all check nodes the sum of the neighboring positions among the message nodes is zero.

The graph representation is analogous to a matrix representation by looking at the adjacency matrix of the graph: let $H$ be a binary $r \times n$-matrix in which the entry $(i, j)$ is 1 if and only if the $i$-th check node is connected to the $j$-th message node in the graph. Then the LDPC code defined by the graph is the set of vectors $c = (c_1, \dots, c_n)$ such that $H \cdot c^{\tp} = 0$. The matrix $H$ is called a \emph{parity check matrix} for the code. Conversely, any binary $r \times n$ matrix gives rise to a bipartite graph between $n$ message and $r$ check nodes, and the code defined as the null space of $H$ is precisely the code associated to this graph. Therefore, any linear code has a representation as a code associated to a bipartite graph (note that this graph is not uniquely defined by the code). However, not every binary linear code has a representation by a \emph{sparse} bipartite graph. If it does, then the code is called a low-density parity-check (LDPC) code.

An important subclass of LDPC codes that feature encoding advantages over other codes of the same class is that of quasi-cyclic low-density parity-check (QC-LDPC) codes\cite{tanner,chen-xu-djurdjevic}. In general, an $[n, k]$-QC-LDPC code satisfies $n = n_0 b$ and $k = k_0 b$ (thus also $r = r_0 b$) for some $b$, $n_0$, $k_0$ (and $r_0$), and admits a parity-check matrix consisting of $n_0 \times r_0$ blocks of $b \times b$ sparse circulant submatrices. Of particular importance is the case where $b = r$ (and hence $r_0 = 1$ and $k_0 = n_0 - 1$), since a systematic parity-check matrix for this code is entirely defined by the first row of each $r \times r$ block. We say that the parity-check matrix is in \textit{circulant form}.

\subsection{QC-MDPC codes}\label{sec:qc-mdpc}

A cryptographically interesting subclass of the QC-LDPC family is that of quasi-cyclic \emph{moderate-density parity-check} (QC-MDPC) codes~\cite{misoczki-sendrier-tillich-barreto}.

QC-MDPC codes in this sense are an entirely distinct class from a family of algebraic codes also known as `MDPC' and designed by Ouzan and Be'ery~\cite{ouzan-be'ery}, despite the name clash. The goal set forth in the latter approach is to obtain high-rate codes of short to moderate length whose duals contain known intermediate-weight words (and thus admit parity-check matrices of moderate density, hence the `MDPC' name), but still have a good error correction capability in comparison with algebraic codes like BCH with similar length and rate. Examples from~\cite{ouzan-be'ery} indicate that typical densities are in the range 17\% to 28\% of the code length. Thus they are indeed intermediate between usual LDPC codes and general ones, but the density is too high for conventional LDPC decoding techniques, to the effect that those codes are not classical Gallager codes in the sense that the density of their dual codes puts them beyond the capability of decoding techniques like plain belief propagation and bit flipping, and especially tailored decoders must thus be adopted. 

By contrast, QC-MDPC codes in the sense of Misoczki \emph{et al.} are oriented toward cryptographic purposes, with densities close enough to LDPC codes as to enable decoding by Gallager's simpler (and arguably more efficient) belief propagation and bit flipping methods, yet dense enough to prevent attacks based on the presence of too sparse words in the dual code like the Stern attack~\cite{stern} and variants, without loosing too much of the error correcting capability so as to keep information-set decoding attacks~\cite{bernstein-lange-peters,bernstein-lange-peters:bcd} infeasible as well. Furthermore, to prevent structural attacks as proposed by Faug\`{e}re \emph{et al.}~\cite{faugere-otmani-perret-tillich} and by Leander and Gauthier~\cite{gauthier-leander}, cryptographically-oriented codes must remain as unstructured as possible except for the hidden trapdoor that enables private decoding and, in the case of quasi-cyclic codes, external symmetries that allow for efficient implementation. Finally, the very circulant symmetry might introduce weaknesses as pointed out by Sendrier~\cite{sendrier-doom}, but these induce only a polynomial (specifically, linear) gain in attack efficiency, and a small adjustment in parameters copes with it entirely. Typical densities in this case are in range 0.4\% to 0.9\% of the code length, one order of magnitude above LDPC codes but well below the published MDPC range above, and certainly within the realm of Gallager codes. Construction is also as random as possible, merely keeping the desired density and circulant geometry, and code lengths are much larger than typical MDPC values.

\subsection{Gallager's Hard Decision (Bit Flipping) Decoding Method}\label{sec:hdd-recap}

We briefly recapitulate Gallager's hard decision decoding algorithm, closely following the very concise and clear description by Huffman and Pless~\cite{huffman-pless}. This will provide the basis for the efficient variant we propose for embedded platforms.

Assume that the codeword is encoded with a binary LDPC code $\mathscr{C}$ for transmission, and the vector $c$ is received. In the computation of the syndrome $s = c H^{\tp}$, each received bit of $c$ affects at most $d_v$ components of that syndrome. If only the $j$-th bit of $c$ contains an error, then the corresponding $d_v$ components $s_i$ of $s$ will equal 1, indicating the parity check equations that are not satisfied. Even if there are some other bits in error among those that contribute to computation of $s_i$, one expects that several of the $d_v$ components of $s$ will equal 1. This is the basis of Gallager's \textit{hard decision decoding}, or bit-flipping, algorithm.

\begin{enumerate}
	\item Compute $c H^{\tp}$ and determine the unsatisfied parity checks (namely, the parity checks where the components of $c H^{\tp}$ equal 1). \label{step:syndrome}
	\item For each of the $n$ bits, compute the number of unsatisfied parity checks involving that bit. \label{step:count-unsat}
	\item Flip the bits of $c$ that are involved in the largest number of unsatisfied parity checks. \label{step:threshold}
	\item Repeat steps \ref{step:syndrome}, \ref{step:count-unsat}, and \ref{step:threshold} until either $c H^{\tp} = 0$, in which case $c$ has been successfully decoded, or until a certain bound in the number of iterations is reached, in which case decoding of the received vector has failed.
\end{enumerate}

The bit-flipping algorithm is not the most powerful decoding method for LDPC codes; indeed, the belief propagation technique~\cite{gallager,huffman-pless} is known to exceed its error correction capability. However, belief-propagation decoders involve computing ever more refined \emph{probabilities} that each bit of the received word $c$ contains an error, thus incurring floating point arithmetic or suitable high-precision approximations and computationally expensive algorithms. In a scenario where the number of errors is fixed and known in advance, as is the case of cryptographic applications, parameters can be designed so that the more powerful but also more complex and expensive belief propagation methods are not necessary for decoding.

We therefore focus on the problem of designing an optimized variant of bit-flipping decoding for highly constrained platforms. Specifically, such methods still suffer from the drawback of requiring a large amount of ancillary memory for counters: if each column of $H$ has Hamming weight $d_v$, step \ref{step:count-unsat} requires $(\lfloor \lg d_v \rfloor + 1)$ bits to store the number of unsatisfied parity-checks for each of the $n$ bits of $c$, hence $n (\lfloor \lg d_v \rfloor + 1)$ bits overall. Besides, steps \ref{step:count-unsat} and \ref{step:threshold} involve a loop of length $n$ each, introducing processing inefficiency. We will show how to avoid these drawbacks in Section~\ref{sec:decoder}.

\subsection{McEliece and Niederreiter encryption}\label{sec:mce-n}

The McEliece encryption scheme was proposed by R. McEliece \cite{mceliece} in 1978. In that scheme, the public key is a generator matrix for a certain code whose decoder is taken to be the private key. An equivalent scheme using a parity-check matrix as public key was proposed by H. Niederreiter in 1986 \cite{niederreiter}. We briefly review these schemes, which consist of three algorithms (\textsf{KeyGen}, \textsf{Encrypt}, \textsf{Decrypt}) each.

\subsubsection{McEliece}

\begin{itemize}
\item \textsf{KeyGen}: Select a binary $t$-error correcting $[n,k]$-code $\mathscr{C}$ with a decoding trapdoor $\mathscr{D}$ and a $k \times n$ generator matrix $G$ in systematic form. The public key is $(G, t)$, and the private key is the decoding trapdoor $\mathscr{D}$.

\item \textsf{Encrypt}: To encrypt a plaintext $m \in \F_2^k$ into a cryptogram $c \in \F_2^n$, select a uniformly random error pattern $e \in \F_2^n$ and weight $t$, and compute $c \gets m \cdot G + e$.

\item \textsf{Decrypt}: To decrypt $c \in \F_2^n$, apply the decoding trapdoor $\mathscr{D}$ to correct the $t$ errors in $c$ (thus finding the error pattern $e \in \F_2^n$ of weight $t$), then extract $m \in \F_2^k$ from the first $k$ columns of $c - e$.
\end{itemize}

\subsubsection{Niederreiter}

\begin{itemize}
\item \textsf{KeyGen}: Select a binary $t$-error correcting $[n,k]$-code $\mathscr{C}$ with a decoding trapdoor $\mathscr{D}$ and an $r \times n$ parity-check matrix $H$ in systematic form, where $r = n - k$. The public key is $(H, t)$, and the private key is the decoding trapdoor $\mathscr{D}$.

\item \textsf{Encrypt}: To encrypt a plaintext $m \in \Z/\binom{n}{t}\Z$ into a cryptogram $c \in \F_2^r$, encode $m$ into a vector $e \in \F_2^n$ of weight $t$ via some conventional permutation unranking method, and compute $c \gets e \cdot H^{\tp}$.

\item \textsf{Decrypt}: To decrypt $c \in \F_2^n$, apply the decoding trapdoor $\mathscr{D}$ to the syndrome $c$ (thus finding the corresponding vector $e \in \F_2^n$ of weight $t$), then decode $m \in \Z/\binom{n}{t}\Z$ from $e$ via permutation ranking.
\end{itemize}

Although the security of these two schemes is equivalent, Niederreiter is the more efficient~\cite{canteaut-sendrier}, being therefore the method of choice for constrained platforms.

\section{An efficient family of MDPC codes}\label{sec:cs-mdpc}

The QC-MDPC codes~\cite{misoczki-sendrier-tillich-barreto} are arguably among the most efficient settings for code-based cryptosystems. However, QC-MDPC parameters for practical security levels, specifically those corresponding to a cost between a legacy-level $2^{80}$ and a top-level $2^{256}$ steps to mount the best possible attacks, yield key and ciphertext space overheads well above the corresponding values achievable with the RSA cryptosystem, which is perhaps the most widely deployed asymmetric cryptography scheme today, and constitutes for that reason a practical upper bound for the corresponding parameters in other cryptographic schemes. Therefore one cannot claim that those codes are generically suitable for constrained platforms.

It turns out we can do better than that with a proper subset of QC-MDPC codes. To define it, we now introduce a class of matrices that admit a space-efficient representation:

\begin{definition}
Given a ring $\R$ and an integer $p$, the set of \emph{cyclosymmetric} matrices of order $p$ over $\R$ is the set of $\Delta_p(\R)$ of square $p \times p$ matrices over $\R$ that are both circulant and symmetric.
\end{definition}

Cyclosymmetric matrices constitute a subring of the ring $\R^{p \times p}$ of $p \times p$ matrices over $\R$, which can be seen by the fact that the identity matrix is cyclosymmetric and that the product of symmetric matrices is itself symmetric iff the factors commute, and indeed circulant matrices are commutative: $(AB)^{\tp} = B^{\tp} A^{\tp} = BA = AB$ (all other properties are trivial). We call this the \emph{cyclosymmetric ring} of order $p$ over $\R$. 

A cyclosymmetric ring can be itself defined over another cyclosymmetric ring and so on recursively, yielding a \emph{multilayered} cyclosymmetric ring ultimately defined over a non-cyclosymmetric ring. This ring tower is written as $\Delta_{p_1} (\dots \Delta_{p_L}(\R_0)$ for successively embedded rings of orders $p_1, \dots, p_L$.
Let $\lyr(\R)$ denote the number of layers of a multilayered cyclosymmetric ring $\R$. We define the number of layers of a non-cyclosymmetric ring $\R_0$ (e.g. a finite field) to be $\lyr(\R_0) = 0$, and then recursively $\lyr(\Delta_p(\R')) = \lyr(\R')+1$. Thus, $\lyr(\Delta_{p_1} (\dots \Delta_{p_L}(\R_0)) = L$.

The interest in a cyclosymmetric ring resides in the fact that elements of $\Delta_p(\R)$ can be represented as a sequence of $\lfloor p/2 \rfloor+1$ elements from $\R$, asymptotically occupying only half the space required by a merely circulant matrix of order $p$ over $\R$. To see this, just note that a circulant $p \times p$ matrix has the form $C_{ij} = c_{(j-i) \bmod p}$ where $c$ is the first row of that matrix, while a symmetric matrix satisfies $C_{ij} = C_{ji}$, thus combining both conditions yields $c_{(j-i) \bmod p} = c_{(i-j) \bmod p}$, which for $i = 0$ (since the first row alone defines the entire matrix) simplifies to $c_j = c_{-j \bmod p}$, or $c_j = c_{p-j}$ for $j > 0$. Therefore, the sequence $c_1, \dots, c_{p-1}$ is a palindrome (and $c_0$ is an arbitrary bit), and only $c_0, \dots, c_{\lfloor p/2 \rfloor + 1}$ are independent.

The space efficiency becomes more noticeable for rings with several layers: an element of $\Delta_{p_1} (\dots \Delta_{p_L}(\R_0)$ is represented as $\prod_{i=1}^L(\lfloor p_i /2 \rfloor + 1)$ elements of $\R'$, roughly a fraction $1/2^L$ of the size of a generic circulant matrix of order $\prod_{i=1}^L{p_i}$ over $\R_0$.

Extending the analogy, we define the family of \emph{cyclosymmetric codes}:

\begin{definition}
A \emph{cyclosymmetric (CS) code} over $\F_q$ is a code which admits a block parity-check matrix whose blocks correspond to elements of a (multiplayerd) cyclosymmetric ring.
\end{definition}

In other words, a cyclosymmetric $[n, n - r]$-code admits an $r \times n$ parity-check matrix $H$ with $r = r_0 p$, $n = n_0 p$, consisting of $r_0 \times n_0$ cyclosymmetric blocks of size $p \times p$ over some smaller ring. The natural advantage of these codes is the compact representation of such parity-check matrices. A particularly efficient case occurs when $r_0 = 1$, that is, $H$ is a simple sequence of $n_0$ cyclosymmetric blocks of size $r \times r$: if $H$ is in systematic form, $r = p_1 \cdots p_L$, and $\R = \Delta_{p_1} (\dots \Delta_{p_L}(\F_q) \dots)$, then $H$ occupies only $(n_0 - 1) \prod_{i=1}^L(\lfloor p_i /2 \rfloor + 1) \lg{q}$ bits.

In cryptographic applications, the natural choice is to adopt binary codes, i.e. $q = 2$, and in particular MDPC codes, due to the simplicity of the decoding algorithm and the greatly reduced parameters that these codes allow for every desired security level. A cyclosymmetric MDPC code is, therefore, a CS-MDPC code. Moreover, in the same cryptographic context we not only propose the use CS-MDPC codes, but also to restrict \emph{error patterns} to the same form as the concatenation of first rows of cyclosymmetric matrices, so that these patterns stand themselves for sequences of cyclosymmetric ring elements, as long as this does not affect the security level.

A disadvantage of a too large number of layers is that, on average, each `1'-bit among the $\prod_{i=1}^L(\lfloor p_i /2 \rfloor + 1)$ independent bits of each block of $H$ represents about $2^L$ `1'-bits in the full $r \times r$ block, rapidly increasing the parity-check matrix density and therefore limiting the error correction capability of bit-flipping and related decoders. However, small values of $L$ (one or two, in some cases possibly even three) yield potentially interesting parameters for cryptographic applications.

\subsection{Security considerations} 

An immediate observation on the structure of cyclosymmetric codes is that one can optimize the Stern attack~\cite{stern,bernstein-lange-peters,engelbert-overbeck-schmidt} and its variants by taking advantage of the form of each row of the parity-check matrix when performing linear algebra operations. Indeed, Stern tries to retrieve a row of low density from the dual code; since the first row consists of one element followed by a palindrome, and the remaining rows are rotated versions thereof, one can in principle reduce in half the overall effort incurred by row manipulations. However, this apparent improvement may turn out to be ineffective: linear algebra operations quickly destroy the palindrome structure within the rows, thwarting the optimization.

Leon's attack~\cite{leon} and related ones do not seem to benefit at all either, because they already ignore part of each row involved in linear algebra operations. Interestingly, a brute force attack would be faster than Leon's against cyclosymmetric codes because it would need to test only $\binom{p/2}{w/2}$ rather than $\binom{p}{w}$ elements, yet for any practical choice of parameters that number remains far above the cost of Stern's or similar attacks. For example, parameters for which the cost of the best known variants of Stern is about $2^{80}$ with block size $p=4800$ and private code density $w=45$, the cost of brute force would be $\binom{p/2}{w/2} \approx 2^{177}$.

Similar observation apply to information-set decoding attacks~\cite{bernstein-lange-peters,bernstein-lange-peters:bcd}: at most, one would expect an improvement by a factor of $2^L$ in the attack cost for $L$-level CS-MDPC codes (recall that, in practice, $L \leqslant 2$).

There seems to be no essential restriction to the value of $p$, although a prudent choice would seem to be to take prime $p$ to avoid the possibility of attacking smaller subcodes. No structural attacks along the lines of Faug\`{e}re \emph{et al.}~\cite{faugere-otmani-perret-tillich} or Leander and Gauthier~\cite{gauthier-leander} seems to apply, since the CS-MDPC trapdoor is of a statistical rather than algebraic nature.

Apart from this, CS-MDPC codes appear to inherit most if not all of the security properties of the superclass of QC-MDPC codes, as indicated in Section~\ref{sec:qc-mdpc}. One consequence of all these considerations is that, to the best of our knowledge, the best attacks against CS-MDPC codes are precisely the best attacks against \emph{generic} QC-MDPC codes.

\section{Scaling the implementation to embedded platforms}\label{sec:scaling}

\subsection{General operation}\label{sec:gen-op}

We use a representation of sparse matrices with a plain arithmetic: both matrix-matrix and matrix-vector products coalesce into (vector-vector) cyclic convolution, for which efficient algorithms like Karatsuba~\cite{karatsuba} and FFT are known. However, simple `textbook' multiplication algorithms, slightly modified so as to operate on circulant matrices represented by their first row alone, are not only more compact, but at least as fast (and often faster) than more advanced counterparts because of the sparsity of the arguments. Indeed, the operations that actually occur in the Niederreiter cryptosystem always involve at least one sparse operand:
\begin{itemize}
    \item inversion of a secret, sparse circulant matrix $H$ yielding a public, dense matrix $K$: this is achieved with a carefully tuned extended Euclidean algorithm (see Section~\ref{sec:euclid}).
    \item Computation of the public syndrome of a sparse error vector. This syndrome is the product of the public, dense parity-check matrix $K$ by the sparse vector $e$.
    \item Computation of the private, decodable syndrome $s^{\tp} = He^{\tp}$ corresponding to a given public syndrome $c^{\tp} = Ke^{\tp}$. This is the product $H_{n_0-1}c^{\tp}$ of the sparse secret matrix $H_{n_0-1}$ by the given dense syndrome $c^{\tp}$, since $K = H_{n_0-1}^{-1} H$ and hence $H_{n_0-1}c^{\tp} = H_{n_0-1}Ke^{\tp} = He^{\tp} = s^{\tp}$.
    \item Additionally, our strategy recovers the decodable syndrome $s$ from a modified but nonzero syndrome $\hat{s}$ after a failed decoding attempt. Such an attempt yields an incorrect error vector $\hat{e}$ of weight not exceeding $\mathrm{HDDMARGIN}(t)$ satisfying the relation $s^{\tp} = \hat{s}^{\tp} + He^{\tp}$. Thus, recovering $s$ involves the product $He^{\tp}$ of a sparse matrix by a sparse vector.
\end{itemize}
Interestingly, the McEliece cryptosystem does involve a product of a dense public generator matrix and a dense random vector in semantically secure constructions like Fujisaki-Okamoto~\cite{fujisaki-okamoto}. This is further reason to adopt the Niederreiter cryptosystem on constrained platforms.

\subsection{Space-efficient convolutional inverse}\label{sec:euclid}

The extended Euclidean algorithm yields a time-efficient method to compute the inverse of a circulant matrix $H = \cir(h)$. The technique consists of mapping the array $h$ (with components $h_j$, $0 \leqslant j < r$) to a polynomial $h(x) = \sum_{0 \leqslant j < r}{h_j x^j} \in \F_2[x]$, computing the modular inverse $h(x)^{-1} \pmod{x^r - 1}$, and mapping $h(x)^{-1}$ back to an array denoted $h^{-1}$ such that $H^{-1} = \cir(h^{-1})$.

An apparently less widely known property of the extended Euclidean algorithm is that it admits a space-efficient implementation as well. In its most usual form, when computing $h^{-1} \bmod m$ the algorithm keeps track of four polynomials $f, g, b, c \in \F_2[x]$ (plus two additional polynomials $u, v \in \F[x]$ that are usually only implicit) related by the constraints $f = bh + um$ and $g = ch + vm$. This suggests a naive implementation requiring up to $4r$ bits of storage for those four polynomials. However, polynomials $f$ and $c$ can actually coexist on the same storage area, and similarly for polynomials $g$ and $b$, as long as $r+2$ bits are available for each of these pairs (totaling $2r+4$ bits) because, at any step of the algorithm execution, it holds that $\deg(f) + \deg(c) \leqslant r$ and $\deg(g) + \deg(b) \leqslant r$. One can easily prove this by Floyd-Hoare logic.

\subsection{A space-efficient decoder}\label{sec:decoder}

The technique of bit-flipping decoding has received a substantial amount of attention in the literature since Gallager's discovery of LDPC codes~\cite{gallager,chang-su-chen-liu,cho-sung,guo-hanzo,ngatched-bossert-fahrner-takawira,wu-ling-jiang-xu-zhao-you,miladinovic-fossorier,ngatched-takawira-bossert,wadayama-nakamura-yagita-funahashi-usami-takumi,zarrinkhat-banihashemi-1,zarrinkhat-banihashemi-2,zhou-cockburn-bates}. However, these are mostly concerned with improving the error correction capability rather than reducing computational resource requirements. Even techniques designed for VLSI like the soft bit-flipping (SBF) technique~\cite{cho-kim-ji-sung,cho-kim-sung}, which might be potential candidates for implementation on the small processors typical of an Internet of Things scenario, turn out to take far more ancillary storage (namely, still $O(n)$ for a code of length $n$) than is typically available on those processors.

It turns out that one can entirely avoid the need for the large storage requirements of a bit-flipping decoder. For cryptographic applications, where the number of introduced errors is fixed and known beforehand, the error correction capability is not the central concern, as long as the desired security level can be attained while fitting the available resources. The variant we propose targets precisely this need. We now describe that variant, together with a rationale for each decision. The full method is summarized as Algorithm~\ref{alg:decoder}.

\begin{itemize}
\item \emph{On-the-fly counter update:} The usual bit-flipping strategy requires two passes over the word variables at each step of the decoding process, namely, a first pass to determine the number of parity errors each variable is involved in (thus keeping an array of counters, one for each variable), and a second pass to tentatively correct the most suspicious variables, which are taken to be those whose parity error count is above a certain threshold. While the second pass could in principle be avoided by adopting a carefully designed data structure singling out the positions that do actually exceed the threshold, not only would maintaining such a structure be considerably expensive, but the approach is not effective for the better part of the decoding process since a large fraction of the variables is expected (and experimentally observed) to be deemed suspicious, to the effect that this whole approach turns out to be easily outperformed by plain counters in both storage requirements and processing time.

We avoid the second pass, the complicated data structure, and even the need to keep an array of counters by counting on-the-fly the number of parity errors each variables is involved in, then deciding immediately whether it has to be tentatively corrected, and modifying the syndrome accordingly.

A consequence of this is that the relation between the actual parity-error counts evaluated on-the-fly and the bit flipping threshold value is likely to change at each such correction, and the decisions that will be taken for variables not yet reached may differ from what they would be if the counters were computed separately. In fact, the parity error threshold becomes known only approximately (unless one took the effort to update it by checking all variables again each time one of them is corrected), but this turns up not to be detrimental to a successful correction of all errors; on the contrary, this is empirically observed to \emph{enhance} the chance of a successful decoding for practical parameters. This can be explained by considering that the number of false positives and false negatives in the error detection heuristic for bits not yet processed is reduced whenever one real error is corrected: in other words, there is a better signal-to-noise ratio in the bit reliability estimation that would be missed if all counters were computed before any actual correction is attempted.

\item \emph{Onset threshold estimation:} As we pointed out, bit flipping works not only with the exact value of the parity-error threshold, but also with a reasonable estimate thereof. This holds equally well at the onset of the process, so that not even the initial parity-error threshold $\theta_0$ needs to be exact.

Analytically deriving a reasonable initial value, however, proves to be rather difficult, but it is easy to bypass this problem by adopting an experimental estimate. This is done by generating a number (say, of order $10^3$) of codes uniformly at random, then performing for each one a number (say, of order $10^3$ as well) of decodings of uniformly generated error patterns of suitable weight, and finally tallying the initial maximum count of parity errors influenced by each variable. The empirical estimate of the initial parity-error threshold $\theta_0$ is then taken to be the average of those maximum counts. The standard deviation is observed to be fairly small, so this approximation, which lies around a fraction 0.7--0.8$\,d_v$ according to the values of $r$, $t$, and $d_v$ itself, leads to a surprisingly stable decoding behavior.

\item \emph{Threshold fine tuning:} The actual parity-error threshold for bit-flipping does not need to be the very maximum current parity-error count among all variables. A faster variant is achieved by setting the threshold somewhere, say $\delta$ parity errors, below that maximum. Experimentally, a fine-tuned $\delta$ can improve decoding speed by an order of magnitude, so this variant is worthwhile.

However, not only the decoding speed, but also the likeliness of decoding failure increases with growing $\delta$, imposing a cutoff at a certain optimal point. As in the case of the initial threshold estimate, deriving an analytical value for the optimum $\delta$ is a difficult and elusive task. We therefore adopt an empirical estimate obtained from simulations here as well.

\item \emph{Decoding failure handling:} Because a large $\delta$ makes a decoding failure more likely, the decoder must be prepared to decrease the actual $\delta$ and restart the process.

Fortunately, rewinding the process to recover the original is easy to do in-place, as the difference between the original syndrome and the current one is the syndrome of the partial error pattern constructed by the decoder up to the failure detection.

Decoding failure is usually detected when a maximum number of decoding attempts is exceeded. Early detection is possible, however, by following the evolution of the weight of error pattern being reconstructed. Although that weight can temporarily surpass the final weight of $t$ errors, in a successful decoding the provisional weight is very unlikely to be too large. A simple and sensible upper limit obtained from simulations is $3t/2$ errors (i.e. allow the decoding process to accumulate spurious errors up to 50\% above the $t$ limit before deciding for failure and decreasing $\delta$), since no successful decoding has been observed to reach as high as this margin before the process begins to reduce it and converge to zero errors.

\item \emph{Simple supporting algorithms:} Sophisticated algorithms with a good asymptotic behavior turn out to be an unnecessary hindrance in the context of decoding at practical cryptographically-oriented parameters.

Thus, for instance, even though convolution-style algorithms may seem ideal to handle products of circulant matrices, in practice one of the factors is usually so sparse that the much simpler approach of just adding together a few rows or columns of the other factor as indicated by the other factor yields a faster outcome (and smaller executable code).

Likewise, representing the error pattern being reconstructed $e$ as an unsorted list of error coordinates yields the most compact representation of $e$ and is very efficient for cryptographic applications because of the relatively small target weight of $e$, even though this incurs sequential searches and updates. 
\end{itemize}

Taking all this into account, we describe in Algorithm~\ref{alg:decoder} an efficient variant of the hard-decision decoding method tailored for platforms with highly constrained data and code storage and processing power.

\begin{algorithm}
\caption{Efficient hard-decision decoder for constrained platforms}\label{alg:decoder}
\begin{algorithmic}[1]
\Input $H \in \F_2^{r \times n}$ (with $n = n_0 r$), a systematic quasi-cyclic low-density parity-check matrix with constant column weight $d_v$, represented as an array of $n_0$ lists of the $d_v$ coordinates of the nonzero components in each cyclic block of $H$.
\Input $s \in \F_2^r$, a bit array representing the received syndrome.
\Input $\delta$, a threshold margin.
\Input $\theta_0$, an estimate of the largest number of unsatisfied parity checks among the $n$ variables (bits) of the codeword with errors.
\Input $iterBound$, a limit on the number of iterations for successful decoding (the heuristic default is $iterBound = t$),
\Output $e \in \F_2^n$, a sparse vector of weight $\wt(e) \leqslant t$ represented as a list of coordinates of its nonzero components (but able to hold the coordinates of $\mathrm{HDDMARGIN}(t) > t$ such coordinates), or $\varnothing$ upon failure.
\Remark compute mod remainders via iterated subtraction.\;

\State $retry \gets \mathbf{false}$
\Repeat
    \State $ew \gets 0$ \ZComment initialize $e$ to no errors
    \State $iter \gets 0$
    \State $\theta \gets \theta_0$ \ZComment initial estimate
    \Repeat
        \LComment Change the bits of the codeword with errors that are involved in the largest number of unsatisfied parity checks:
        \State $newmax \gets 0$ \ZComment new estimate for $\theta$
        \For{$j \gets 0 \To n - 1$}
            \State $L \gets H[\lfloor j / r \rfloor]$
            \State $unsat \gets 0$
            \For{$z \gets 0 \To d_v - 1$}
                \If{$s[(j + L[z]) \bmod r] = 1$}
                    \State $unsat \gets unsat + 1$
                \EndIf
            \EndFor
            \State $newmax \gets \max\{unsat, newmax\}$
            \If{$unsat \geqslant \theta - \delta$} \ZComment try to correct: 
                \If{$\exists q \in [0..ew-1]$ {\SuchThat} $e[q] = j$}
                    \State $ew \gets ew - 1, \; e[q] \gets e[ew]$
                \ElsIf{$ew < \mathrm{HDDMARGIN}(t)$}
                    \State $e[ew] \gets j, \; ew \gets ew + 1$
                \Else{} \ZComment too many spurious errors introduced 
                    \State \textbf{break} \ZComment to line \ref{stmt:iter}
                \EndIf
                \For{$z \gets 0 \To d_v - 1$} \ZComment update syndrome:
                    \State $i \gets (j + L[z]) \bmod r$
                    \State $s[i] \gets \neg s[i]$
                \EndFor
            \EndIf
        \EndFor
        \State\label{stmt:iter} $\theta \gets newmax$
        \LComment Iterate until the syndrome is zero (or until a bound on the number of iterations is reached)
        \State $iter \gets iter + 1$
    \Until $\wt(s) = 0 \Or iter = iterBound$

\algstore{bkbreak}
\end{algorithmic}
\end{algorithm}
\addtocounter{algorithm}{-1}
\begin{algorithm}
\caption{(Continued)}
\begin{algorithmic}[1]
\algrestore{bkbreak}

    \If{$(\wt(s) \neq 0 \Or ew > t) \And \delta > 0$}
        \State $\delta \gets \delta - 1$ \ZComment threshold margin was too high
        \For{$q \gets 0 \To ew - 1$} \ZComment revert syndrome to original form:
            \State $j \gets e[q]$
            \State $L \gets H[\lfloor j / r \rfloor]$
            \For{$z \gets 0 \To d_v-1$}
                \State $i \gets (j + L[z]) \bmod r$
                \State $s[i] \gets \neg s[i]$ 
            \EndFor
        \EndFor
        \State $retry \gets \mathbf{true}$
    \EndIf
\Until $\textbf{not} \; retry$

    \If{$\wt(s) = 0 \And ew \leqslant t$}
        \State \Return{$e, ew$}
    \Else
        \State \Return{$\varnothing$}
    \EndIf

\end{algorithmic}
\end{algorithm}

\section{Suggested parameters}\label{sec:params}

For the sake of illustration, sample CS-MDPC parameters for typical security levels are listed on Tables~\ref{tab:pic-params1} and~\ref{tab:pic-params2}.

\begin{table}[htp]\centering
\caption{CS-MDPC parameters for Niederreiter encryption (1 layer; $n_0 = 2$)}\label{tab:pic-params1}
\begin{tabular}{ccccccc}\hline
$r$                       & $|pk|$ (bits)           & $d_v$ & $t$ & $\theta_0$ & $\delta$ & sec\\
$4801$                    & $2401$                  & 45    & 84  & 37         & 9        & $2^{80}$\\
$7839$                    & $3919$                  & 65    & 117 & 48         & 4        & $2^{112}$\\
$9863$                    & $4931$                  & 71    & 134 & 55         & 5        & $2^{128}$\\
$20487$                   & $10243$                 & 105   & 198 & 75         & 8        & $2^{192}$\\
$32771$                   & $16386$                 & 137   & 264 & 105        & 10       & $2^{256}$\\
\hline
\end{tabular}
\end{table}

\begin{table}[htp]\centering
\caption{CS-MDPC parameters for Niederreiter encryption (2 layers; $n_0 = 2$)}\label{tab:pic-params2}
\begin{tabular}{ccccccc}\hline
$r$                       & $|pk|$ (bits)           & $d_v$ & $t$ & $\theta_0$ & $\delta$ & sec\\
$61\!\times\!79= 4819$    & $31\!\times\!40= 1240$  & 45    & 84  & 37        & 9         & $2^{80}$\\
$47\!\times\!167= 7849$   & $24\!\times\!84= 2016$  & 65    & 117 & 48        & 4         & $2^{112}$\\
$71\!\times\!139= 9869$   & $36\!\times\!70= 2520$  & 71    & 134 & 55        & 5         & $2^{128}$\\
$103\!\times\!199= 20497$ & $52\!\times\!100= 5200$ & 105   & 198 & 75        & 8         & $2^{192}$\\
$73\!\times\!449= 32777$  & $37\!\times\!225= 8325$ & 137   & 264 & 105       & 10        & $2^{256}$\\
\hline
\end{tabular}
\end{table}

Although key sizes still fall short of reaching typical values for pre-quantum elliptic curve cryptosystems, CS-MDPC Niederreiter encryption keys become competitive with pre-quantum RSA and post-quantum, size-optimal NTRU for non-legacy security levels, namely, $2^{112}$ onward. We also compare the key sizes with the previous smallest code-based parameters, namely, those attainable with QC-MDPC codes. This can be shown on Table~\ref{tab:compare}. Besides, as we will see in Section~\ref{sec:experim}, the result is still competitive with elliptic curve implementations on constrained platforms according to other relevant metrics.
\begin{table}[htp]\centering
\caption{Public key and cryptogram size comparison (sizes in bits)}\label{tab:compare}
\begin{tabular}{ccccc}\hline
CS-MDPC & RSA     & NTRU  & QC-MDPC & sec\\
$2016$  & 2048    & 4411  & 7836    & $2^{112}$\\
$2520$  & 3072    & 4939  & 9856    & $2^{128}$\\
$5200$  & 7680    & 7447  & 20480   & $2^{192}$\\
$8325$  & 15360   & 11957 & 32768   & $2^{256}$\\
\hline
\end{tabular}
\end{table}

\section{Experimental results}\label{sec:experim}

We assessed the effectiveness of the techniques described herein according to the metrics of ROM and RAM usage by implementing the Niederreiter cryptosystem with the proposed parameters and decoder on the PIC24FJ32GA002-I/SP (32MHz) platform in the C programming language. No assembly language optimization has been attempted.

Mapping from raw plaintext (bit sequences) and error patterns is most efficiently achieved (in processing speed, data storage and executable code size requirements) with the Sendrier technique~\cite{sendrier}. It was natural to adopt the same technique choosing CS-MDPC codes uniformly at random.

The observed program size (i.e. the ROM requirements for the deployed system) with the compiler employed is about 5.8 KiB.
Storage (RAM) requirements are about 2.2 KiB overall, including the space needed for indices, counters and runtime bookkeeping (return addresses, stack management). By contrast, a plain implementation of the bit flipping technique would take at least 7.2 KiB for the counters alone, far above the 3.8 KiB RAM available on a PIC24FJ32GA002 microcontroller.
%
%
For simplicity, we limited the experiments to 1-layer CS-MDPC codes at the 80-bit security level.

In comparison, elliptic curve ElGamal encryption at the same security level on the ATMega128L platform using the state-of-the-art RELIC library~\cite{relic-toolkit} demands over 31 KiB ROM and 2.1 KiB RAM. 

\section{Conclusion}\label{sec:conclusion}

We described how to scale code-based cryptosystems to platforms with very constrained storage and processing resources. Central to our proposal is the adoption of quasi-cyclic LDPC codes coupled with a storage-efficient algorithm for key pair generation, a carefully tailored variant of hard-decision decoding, and fine-tuned parameters. The efficiency of the result is competitive with traditional cryptosystems like those based on elliptic curves.


\bibliographystyle{IEEEtranS}
\bibliography{EmbeddedCS-MDPC}

\end{document}